\documentclass[a4paper, 12 pt, twoside, reqno]{amsart}
     \usepackage[utf8]{inputenc}
    \usepackage{slashed}
    \usepackage[english]{babel}
    \usepackage[all]{xy}
    \usepackage{enumerate}
    \usepackage{setspace}
    \usepackage{hyperref}
    \usepackage{tikz}
\usetikzlibrary{calc}
\usepackage{tikz-cd}
    
    \usepackage[euler-digits,euler-hat-accent]{eulervm}
    
    \usepackage{amsthm}
    
    \usepackage{times}
    \usepackage{amsmath}
    \usepackage{amsfonts}
    \usepackage{amssymb}
    \usepackage{amsthm}
    \usepackage{graphicx}
    \usepackage{array}
    \usepackage{color}
    \usepackage{mathrsfs}
    
    \usepackage{eucal}
    \usepackage{esint} 
    \usepackage{tikz}
    \usepackage{upgreek}
    
    \usepackage{mathtools}

    \allowdisplaybreaks
    
    \theoremstyle{plain}
    \newtheorem{theorem}{Theorem}[section]

    \newtheorem{proposition}[theorem]{Proposition}
    \newtheorem{lemma}[theorem]{Lemma}
    
    \theoremstyle{definition}
    \newtheorem{definition}[theorem]{Definition}

    \theoremstyle{remark}
    \newtheorem{remark}[theorem]{Remark}

    \numberwithin{equation}{section}
    \numberwithin{figure}{section}
    \numberwithin{table}{section}
\newcommand{\W}{\mathcal W}
    \newcommand{\Om}{{\bf \Omega}}

    \newcommand{\R}{\mathbb{R}}
    
    \newcommand{\C}{\mathbb{C}} 
    
    \newcommand{\Z}{\mathbb{Z}}
    \newcommand{\T}{\mathbb{T}}

    \newcommand{\G}{\mathcal{G}}
    \newcommand{\expo}[1]{\,\mathrm{e}^{#1}\,}                 
    \newcommand{ \ii}{\,\mathrm{i}\,}

                    \newcommand{\ie}{\textsl{i.\,e.\,}}

    \newcommand{\A}{\mathfrak{A}}

    \hypersetup{
    pdftoolbar=true,        
    pdfmenubar=true,        
    pdffitwindow=true,     
    pdfstartview=true,    
    pdftitle={Magnetic interfaces models},    
    pdfauthor={Danilo Polo Ojito},     
    breaklinks=true, 
    colorlinks=true,       
    linkcolor=purple,         
    citecolor=teal, 
    urlcolor=blue, 
    bookmarksopen=true, 
    filecolor=magenta,      
    }

\begin{document}

    \title[Topological Kirchhoff Laws for magnetic interfaces]{Topological Kirchhoff Laws for systems with magnetic interfaces}

\author[D. P. Ojito]{Danilo Polo Ojito}
\address[D. P. Ojito]{Department of Mathematics, Pontificia Universidad Javeriana, Bogota, Colombia}
	\email{danilopolo@javeriana.edu.co}

    \vspace{2mm}
    
    \date{\today}

    \dedicatory{Dedicated to Jean Bellissard on the occasion of his
80th birthday}
    \maketitle

    \begin{abstract}
We consider a finite rectangular cell complex in the plane whose faces are occupied by quantum systems subject to constant and possibly different
magnetic fields. Adjacent bulk phases determine magnetic
interfaces along the edges of the graph, while its vertices describe junctions at which several interfaces meet. Using K-theory for operator algebras, we show that the current along each oriented edge is determined by the
difference of the bulk Chern numbers of the adjacent phases. As a consequence, the oriented sum of the interface currents meeting at every vertex vanishes. This provides a topological description of the  Kirchhoff law for
magnetic interface currents on finite planar graphs.

\medskip

    \end{abstract}

    \tableofcontents

\section{Introduction}
In this work, we study topological currents flowing along interfaces
between two-dimensional systems, each subject to a constant magnetic field. Such currents may arise when the insulating phases on either
side of an interface are topologically distinct. The integer quantum
Hall effect~\cite{vonKlitzing1980} provides a prototypical example of
electronic transport along a boundary. In a strong perpendicular
magnetic field, a two-dimensional electron gas supports current-carrying
edge modes~\cite{Halperin1982}. Their net chirality is stable under
local perturbations and reflects the bulk Chern number that determines the quantized Hall conductance~\cite{TKNN1982}. A similar mechanism operates at the interface between two insulators
with a common bulk gap. When their bulk Chern numbers differ, modes
localized near the interface carry a net current whose direction is
determined by the sign of that difference, \ie  the relative Chern number. These phenomena illustrate the bulk--edge and bulk--interface correspondences, respectively, according to which the dynamics of the electrons along the edge or interface is  fully determined by an input coming from the bulk.

\medskip 

The pioneering $C^*$-algebraic approach to solid-state physics developed by
Bellissard and coworkers~\cite{Bel1,Bel,SBB1998} provides a natural framework
for understanding the quantization of transport coefficients through
$K$-theory and non-commutative geometry. Since this approach does not
require lattice periodicity, it also applies to disordered and amorphous
materials, for which conventional Bloch band theory is not directly
applicable.  Building on this approach, Kellendonk, Schulz-Baldes, and
collaborators~\cite{Kel,Kel1, PSbook} formulated the bulk--edge correspondence in terms of the short exact sequence of $C^*$-algebras
$$
0 \longrightarrow \A_{\mathrm{edge}}
  \longrightarrow \A_{\mathrm{half}}
  \longrightarrow \A_{\mathrm{bulk}}
  \longrightarrow 0,$$
where $\A_{\rm bulk}$ describes the bulk observables, $\A_{\rm edge}$ describes the observables localized near the boundary, and $\A_{\rm half}$ describes the full system, including its boundary. Moreover,  the quotient map sends an observable to its bulk counterpart and its kernel consists of observables localized near 
edge. When the Fermi energy $\mu$ lies in a bulk spectral gap of the Hamiltonian $H$, the associated
boundary map in $K$-theory sends the class  $[P_\mu]=[\chi_{\leq \mu}(H)]$ of the bulk Fermi projection
in $K_0(\A_{\mathrm{bulk}})$ to a class in
$K_1(\A_{\mathrm{edge}})$. Pairing these classes with suitable Chern cocycles, identifies the bulk Chern number with the edge winding number for compatible
orientations. Consequently, a bulk invariant determines the quantized
edge conductance~\cite{BKR2017,Kel,Kel1,PSbook}. A similar approach applies to an interface separating two bulk phases,
where the difference of their Chern numbers determines the quantized
interface current~\cite{Bourne2026, Kot,DG2021,DGO2026,Ojito2023}.

\medskip

A concrete realization of this bulk--interface correspondence arises
when the magnetic field varies across a single two-dimensional
material. In the step-field version of the Iwatsuka model~\cite{Iwa},
the field takes different constant values on the two sides of a
straight line, which thus defines a magnetic interface. From a
mathematical perspective, the existence and quantization of currents
along such interfaces have been studied in both continuum and lattice
models~\cite{Dom,His,DG2021,DGO2026,Ojito2023}. In this work, we extend the geometry of the Iwatsuka model from a single straight interface to a network of magnetic interfaces within a single two-dimensional material. We consider a finite rectangular grid in the plane and denote the resulting cell complex by $\mathfrak G=(F,E,V)$, where $F$, $E$, and $V$ are its sets of faces, edges, and vertices, respectively. The magnetic field is constant on each face and is therefore specified by a function $B\colon F\to\R$. Each edge separates two adjacent faces and defines a magnetic interface when their field values differ. Orienting each edge distinguishes the face on its left from the face on its right. At each vertex, where several edges meet, these orientations also distinguish incoming from outgoing edges (see Figure~\ref{fig: graph}).

Following the approach of~\cite{DG2021}, we construct a magnetic $C^*$-algebra $\A_B$ that describes the full system and encodes the incidence structure of the graph $\mathfrak G$, recording which faces share an edge and which edges meet at a vertex. This structure is reflected in a filtration of $\A_B$ by ideals corresponding to the stratification of $\mathfrak G$ by dimension. The associated subquotients are the face, edge, and vertex algebras $\A_F$, $\A_E$, and $\A_V$, which describe the bulk-face regions, the magnetic interfaces, and their junctions, respectively. The relations among the corresponding face, edge, and vertex algebras
are encoded by two boundary maps in $K$-theory,
$$
K_0(\A_F)
  \xrightarrow{\;\partial_0^{\mathrm{FE}}\;}
K_1(\A_E)
  \xrightarrow{\;\partial_1^{\mathrm{EV}}\;}
K_0(\A_V).$$
To relate the boundary maps to transport, let $\widehat H\in\A_B$ be a
self-adjoint Hamiltonian and, for each face $f$, let $H_f$ denote
its image in the corresponding bulk algebra. Suppose that the
Hamiltonians $H=(H_f)_{f\in F}$ have a common spectral gap $\Delta$. For a Fermi
energy $\mu\in\Delta$, the Fermi projection
$P_{\mu,f}=\chi_{(-\infty,\mu]}(H_f)$ defines the bulk Chern number
$\operatorname{Ch}(P_{\mu,f})$ of the face $f$. The first boundary map $\partial_0^{\mathrm{FE}}$ records, at each
oriented edge, the signed difference between the Chern numbers of
the two adjacent faces. Our main Theorem identifies this difference,
up to a universal factor, with the topological current along the
edge. The second boundary map $\partial_1^{\mathrm{EV}}$ takes the
signed sum of the edge winding numbers at each vertex. The identity
$\partial_1^{\mathrm{EV}}\circ\partial_0^{\mathrm{FE}}=0$ thus
expresses Kirchhoff's current law: at every junction, the total
incoming current equals the total outgoing current.

\begin{theorem}
Let $\Delta$ be an interval contained in a common spectral gap of the
bulk Hamiltonians $H=(H_f)_{f\in F}$, and let $\mu\in\Delta$.  For every oriented edge $e\in E$, the current density $J_e$
along $e$ is given by the formula
$$
    J_e\;
    =\;\frac{\mathfrak e}{2\pi\hbar}
      \sum_{\{f\in F\,\mid\,e\in\partial f\}}
      \operatorname{sign}(f,e)\,\operatorname{Ch}(P_{\mu,f}),$$
where $\mathfrak e$ denotes the electron charge, $\hbar$ is the reduced Planck constant, and $\operatorname{sign}(f,e)$ is the oriented incidence number of $e$ in the boundary of $f$. Moreover, Kirchhoff's current law holds at every vertex $v\in V$, \ie 
$$
    \sum_{\{e\in E\,\mid\,\partial_1(e)=v\}} J_e
    \;=\;
    \sum_{\{e\in E\,\mid\,\partial_0(e)=v\}} J_e,
$$
where $\partial_0(e)$ and $\partial_1(e)$ denote the initial and
terminal vertices of $e$, respectively.
\end{theorem}

\medskip
\noindent
\textbf{Acknowledgments.} Danilo Polo Ojito thanks Giuseppe De Nittis for introducing him to this topic and for many stimulating discussions. This work was supported by the
U.S. National Science Foundation through grant CMMI-2131760
and by the U.S. Army Research Office through contract
W911NF-23-1-0127.

\medskip
\noindent
\textbf{AI statement.} The author used ChatGPT 6.2 ASTRA to assist
with drafting and editing the manuscript. The author reviewed the entire
manuscript and takes full responsibility for its content.

\section{Magnetic interfaces and their junctions}
\subsection{The setting}
\begin{figure}
\centering
\begin{tikzpicture}[scale=0.8]
  \foreach \y in {0,1.5,3}
    \draw (-0.5, \y) -- (6.5, \y);

  \foreach \x in {1,3,5}
    \draw (\x,-1) -- (\x,4);

  \foreach \x in {1,3,5}
    \foreach \y in {0,1.5,3}
      \fill[blue] (\x,\y) circle (2.5pt);

  \node at (3.2,1.7) {\textcolor{red}{$v$}};
  \node at (2,2.2) {$B(f_v^2)$};
    \node at (4,2.2) {$B(f_v^1)$};
      \node at (2,0.7) {$B(f_v^3)$};
    \node at (4,0.7) {$B(f_v^4)$};
 
\end{tikzpicture}\;\qquad\; 
\caption{A rectangular decomposition of the plane. The labels indicate the magnetic field values on the four faces adjacent to the vertex $v$.} \label{fig: graph}
\end{figure}
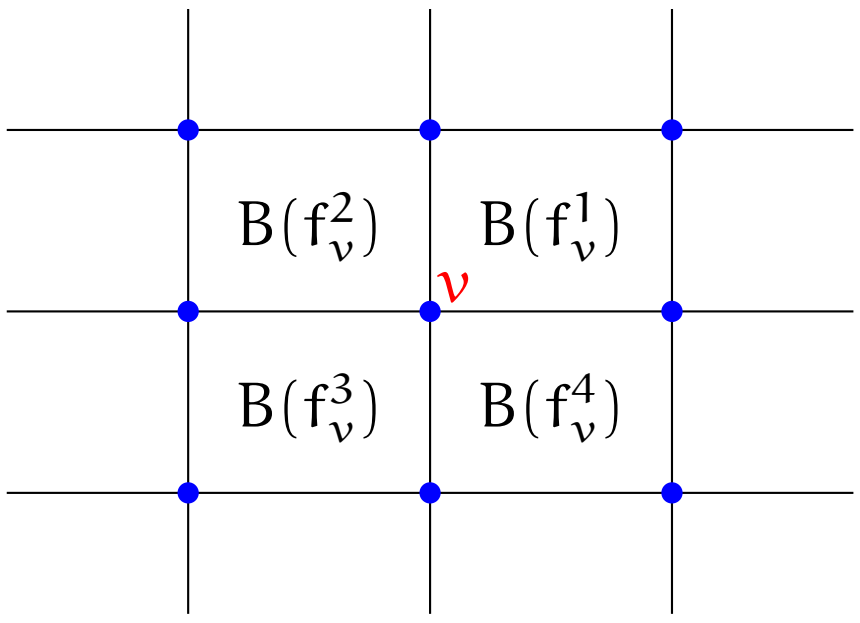

We consider a network of two-dimensional bulk systems,
each subject to a constant magnetic field, joined along
magnetic interfaces. These interfaces separate neighboring
bulk regions and meet at junctions. To simplify the exposition,
we arrange the bulk regions in a rectangular decomposition
of the plane, as illustrated in Figure~\ref{fig: graph}.
More general arrangements, including junctions with different opening angles, could be investigated
within the same framework together with \cite{DGO2026,OPS2}.

Let $\mathfrak G=(V,E,F)$ denote the rectangular decomposition
associated with the network,
whose faces $F$, edges $E$, and vertices $V$ represent the bulk regions,
magnetic interfaces, and their junctions, respectively.
We denote by $\partial f$ the set of edges bounding a face $f$,
and by $\partial_0(e)$ and $\partial_1(e)$ the initial and final
vertices of an edge $e$, respectively.
We orient horizontal edges from left to right and vertical
edges from bottom to top, and assign the clockwise orientation
to each face. For $e\in\partial f$, we set ${\rm sign}(f,e)=1$
if the orientation of $e$ agrees with the boundary orientation
induced by $f$, and ${\rm sign}(f,e)=-1$ otherwise.

\medskip

 The magnetic fields of the bulk systems are specified by a
function $B\colon F\to\R$, where $B(f)$ is the constant field
in the region associated with the face $f$. Since the set of field
values $\mathcal B:=\{B(f)\mid f\in F\}$ is finite, the
configuration space $\Om:=\mathcal B^{\Z^2}$ is compact
and Hausdorff in the product topology.  In particular, if
$|\mathcal B|>1$, it is a Cantor space. The group $\mathbb Z^2$ acts continuously on  $\Om$ via translations:
$$
[\tau_m(\omega)](n)\;:=\;\omega(n+m),
\qquad m,n\in\mathbb Z^2,\;\omega\in {\bf \Omega}$$
From now on we assume that distinct faces carry distinct magnetic fluxes
modulo $2\pi$, that is,
$B(f)-B(f')\notin 2\pi\Z$ for any $f\neq f'$. 

\begin{remark}
   The previous assumption ensures that distinct faces determine different bulk configurations and that every edge in $E$ represents a physical magnetic
interface. The network is therefore completely determined by the data $(\mathfrak G,B)$. The case of equal fluxes on adjacent faces allows their common interface
to be removed, yielding a subgraph of $\mathfrak G$;
repeated fluxes elsewhere may further identify bulk configurations.   \hfill $\blacktriangleleft$
 \end{remark}

 For each vertex $v\in V$, denote the four incident faces by
$f_v^j$, $j=1,2,3,4$, starting with the face in the first quadrant
and proceeding counterclockwise, see figure \ref{fig: graph}.
For $v\in V$ let us define the configuration $b_v\in {\bf \Omega}$ as the function $b_v\colon \Z^2\to \mathcal{B}$ determined by the four faces incident
to $v$, \ie 
$$b_v(n_1,n_2)\;=\;\begin{cases} B(f^1_v), &  n_1,n_2\geq 0\\
B(f^2_v), & n_2\geq0,\; n_1<0\\
B(f^3_v),& n_1,n_2<0\\
B(f^4_v), & n_1\geq 0, \;n_2<0
\end{cases}$$
The configuration $b_v$ describes the junction of the four
bulk systems with the constant magnetic fields specified above,
and thus encodes the local structure of $(\mathfrak G,B)$
at the vertex $v$. Its translates and their limits capture
the corresponding junction, interface, and homogeneous bulk
configurations. These form its \emph{magnetic hull}, defined by
$$
\Xi_v
\;:=\;
\overline{\mathscr{O}(b_v)}
\;\subset\;{\bf \Omega},$$
where $\mathscr{O}(b_v)=\{\tau_n(b_v)\mid n\in \Z^2\}$ is the $\Z^2$-orbit of $b_v$. By construction, this is a closed $\Z^2$-invariant subset of $\Om$. The total magnetic hull of the graph consists of all local magnetic field configurations satisfying the compatibility conditions on overlapping regions. Following the approach in \cite{OPSCMP2025, OPS1}, this is described by the union
\begin{equation}
    \Xi_B
\;:=\;
\bigcup_{v\in V}\Xi_v\;\subset \;{\bf \Omega}
\end{equation}
Since $V$ is finite, $\Xi_B$ is a finite union of compact
$\mathbb Z^2$-invariant subsets of ${\bf\Omega}$. Hence $\Xi_B$
is itself compact, closed, and $\mathbb Z^2$-invariant.

\medskip

For each $f\in F$, define the constant configuration $b_f\in\Om$ by $
    b_f(n)=B(f)$ for all $n\in \Z^2.$ This configuration describes the homogeneous bulk system associated
with the face $f$, subject to the constant magnetic field $B(f)$. For each vertical edge $e$, consider the interface configuration
$b_e\in\Om$ by
$$
    b_e(n_1,n_2)
    \;:=\;
    \begin{cases}
        B(f_e^R), & n_1\geq 0,\\
        B(f_e^L), & n_1<0,
    \end{cases}$$
where $f_e^R$ and $f_e^L$ denote the faces lying to the right and
left of $e$, respectively. Similarly, for each horizontal edge $e$,
set
$$
    b_e(n_1,n_2)
    \;:=\;
    \begin{cases}
        B(f_e^U), & n_2\geq 0,\\
        B(f_e^D), & n_2<0,
    \end{cases}$$
where $f_e^U$ and $f_e^D$ denote the faces lying above and below
$e$, respectively. Thus, $b_e$ describes a straight interface separating two
homogeneous bulk regions with the magnetic fields of the faces
adjacent to $e$. It is invariant under translations parallel
to the interface, while translations in the transverse direction
recover the adjacent constant configurations $b_f$ as limits.
In this way, $b_e$ encodes both the local interface geometry
and its neighboring bulk configurations.
 Let us define the sets
 \begin{equation}
     \mathscr{O}_F\;=\; \bigsqcup_{f\in F}\mathscr{O}(b_f),  \quad    \mathscr{O}_V\;=\; \bigsqcup_{v\in V}\mathscr{O}(b_v), \quad    \mathscr{O}_E\;=\; \bigsqcup_{e\in E}\mathscr{O}(b_e), 
 \end{equation}
These invariant sets provide a convenient description of the configurations of the system.
\begin{proposition}\label{prop: desciprtion of the hull}
The magnetic hull admits the decomposition
$$
\Xi_B\;=\;\mathscr{O}_F\sqcup\mathscr{O}_E\sqcup\mathscr{O}_V
$$
into invariant, locally closed subsets. Moreover, it admits
a filtration by closed invariant subsets
$$
\Xi_0\;\subset\;\Xi_1\;\subset\;\Xi_2=\Xi_B,
$$
where
$
\Xi_0=\mathscr{O}_F$ and  $
\Xi_1=\mathscr{O}_F\sqcup\mathscr{O}_E.$
In particular, $\mathscr{O}_E$ is open in $\Xi_1$,
and $\mathscr{O}_V$ is open in $\Xi_B$.
\end{proposition}
\begin{proof}
By construction, $b_v\in\Xi_B$ for every $v\in V$.
Any edge $e$ is incident to some vertex $v$. The sequence of $\Z^2$-translations of $b_v$ along the direction of $e$ and towards its interior yields $b_e$ as a limit and consequently $b_e\in \Xi_B$.
Likewise, translating $b_e$ in the direction normal to $e$
and towards either adjacent face yields the corresponding
constant configuration $b_f$.
Since $\Xi_B$ is closed and invariant, it follows that
$$
\mathscr{O}_F\sqcup\mathscr{O}_E\sqcup\mathscr{O}_V
\;\subset\;\Xi_B.
$$

For the reverse inclusion, let $\xi\in\Xi_B$. There exist $v\in V$ and a sequence
$n(k)=(n_1(k),n_2(k))\in\Z^2$ such that $
\tau_{n(k)}(b_v)\to \xi.$ After passing to a subsequence, we may assume that each coordinate $n_i(k)$
is either constant or tends to $+\infty$ or $-\infty$.
We distinguish three cases. If both coordinates are constant, then $n(k)=n$ along this
subsequence for some $n\in\Z^2$, and hence
$\xi=\tau_n(b_v)\in\mathscr{O}_V$. If exactly one coordinate is constant and the other tends
to infinity in absolute value, one of the two interfaces of $b_v$ escapes to infinity. The remaining interface determines an edge $e$ incident
to $v$, and the limit is a translate of $b_e$. Thus $\xi\in\mathscr{O}_E$. Finally, if both coordinates tend to infinity in absolute
value, both interfaces escape every finite subset of $\Z^2$.
On each finite subset, the translated configuration is
eventually constant and equal to $B(f)$ for one of the faces
$f$ incident to $v$. Consequently,
$\xi=b_f\in\mathscr{O}_F$. This concludes the reverse inclusion and $\Xi_B=\mathscr{O}_F\sqcup\mathscr{O}_E\sqcup\mathscr{O}_V.$

It remains to establish the closedness assertions.
Each bulk configuration $b_f$ is $\Z^2$-invariant, so
$\Xi_0=\bigsqcup_f \{b_f\}$ is a finite set and is therefore closed.
For each edge $e$, the same analysis of translation limits
gives
$$
\overline{\mathscr{O}(b_e)}
\;=\;
\mathscr{O}(b_e)
\;\sqcup\!
\bigsqcup_{\substack{f\in F\\ e\in \partial f}}
\mathscr{O}(b_f).
$$
Consequently,
$$
\Xi_1
\;=\;
\Xi_0\;\cup\!
\bigcup_{e\in E}\overline{\mathscr{O}(b_e)}
$$
is a finite union of closed subsets of $\Xi_B$ and is
therefore closed. Both $\Xi_0$ and $\Xi_1$ are invariant,
being unions of translation orbits. Finally,
\begin{equation}\label{eq: orbit sets}
    \mathscr{O}_F\;=\;\Xi_0,\qquad
\mathscr{O}_E\;=\;\Xi_1\setminus\Xi_0,\qquad
\mathscr{O}_V\;=\;\Xi_B\setminus\Xi_1.
\end{equation}
Thus all three strata are locally closed,
$\mathscr{O}_E$ is open in $\Xi_1$, and
$\mathscr{O}_V$ is open in $\Xi_B$.
\end{proof}
The Dirac measures $\mu_f:=\delta_{b_f}$, $f\in F$, are
precisely the ergodic invariant probability measures on $\Xi_B$.
Indeed, invariance forces all points in a given translation
orbit to have the same mass. Since the edge and vertex orbits
are countably infinite, every invariant probability measure
assigns zero mass to these orbits and is therefore supported
on the finite set $\mathscr{O}_F$. The translation action also yields the natural identifications
$$
\mathscr{O}(b_e)\;\simeq\;\Z,
\qquad
\mathscr{O}(b_v)\;\simeq\;\Z^2.
$$
Let $\mu_e$ and $\mu_v$ denote the counting measures on these orbits, respectively. Each is invariant and, up to
a nonnegative scalar multiple, is the unique positive invariant
Radon measure on its orbit. Let $\mathfrak M$ be the cone of all positive
$\Z^2$-invariant Borel measures on $\Xi_B$ whose restrictions
to the orbit sets $\mathscr{O}_K$, given in \eqref{eq: orbit sets}, are Radon measures, for $K=F,E,V$. One gets an explicit description of this space:

\begin{proposition}\label{prop: measures}
Every $\mu\in\mathfrak M$ admits a unique decomposition
$$
\mu
\;=\;
\sum_{f\in F}a_f\mu_f
+
\sum_{e\in E}a_e\mu_e
+
\sum_{v\in V}a_v\mu_v,
\qquad a_f,a_e,a_v\geq0.
$$
Consequently, $\mathfrak M$ is naturally isomorphic to
$\R_{\geq0}^{\,|F|+|E|+|V|}$ as a convex cone.
\end{proposition}

\begin{proof}
Let $\mu\in\mathfrak M$. Invariance implies that all points
in the same translation orbit have equal mass.
Moreover, this mass is finite, since each singleton is
compact in its corresponding stratum and the restriction
of $\mu$ to that stratum is Radon. Set
$
a_k:=\mu(\{b_k\})$ for $k=f,e,v.$ Since by Proposition \ref{prop: desciprtion of the hull} $\Xi_B$ is a countable disjoint union of the face,
edge, and vertex orbits, countable additivity gives
$$
\mu
\;=\;
\sum_{f\in F}a_f\mu_f
+
\sum_{e\in E}a_e\mu_e
+
\sum_{v\in V}a_v\mu_v.
$$
Conversely, every such nonnegative combination is invariant
and its restriction to each stratum is Radon.
Finally, evaluating on the singletons $\{b_f\}$, $\{b_e\}$,
and $\{b_v\}$ proves uniqueness.
\end{proof}

\subsection{The groupoid and its cocycle} Let us denote by $\G_B:=\Xi_B\rtimes_\tau \Z^2$ the corresponding \emph{transformation groupoid}. Its unit space is naturally identified with
$\Xi_B$, and its arrows are pairs $(\xi,n)\in\Xi_B\times\Z^2$. We equip $\G_B$ with the product topology, where $\Z^2$ is
discrete. Since $\Xi_B$ is compact Hausdorff and $\tau$ acts by
homeomorphisms, $\G_B$ is a locally compact Hausdorff étale groupoid. 
The range and source maps  $\mathfrak{r,s}\colon \G_B\to \Xi_B$ are
$$
\mathfrak r(\xi,n)\;=\;\xi,
\qquad
\mathfrak s(\xi,n)\;=\;\tau_{-n}(\xi).$$
Consequently, $(\xi,n)$ and $(\eta,m)$ are composable precisely when
$\eta=\tau_{-n}(\xi)$. In that case, their product is
$$
(\xi,n)(\tau_{-n}(\xi),m)\;=\;(\xi,n+m),$$
and the inverse of $(\xi,n)$ is $
(\xi,n)^{-1}=(\tau_{-n}(\xi),-n).$ The space of composable pairs is denoted by $\G_B^{(2)}$. Since $\G_B^{(2)}\simeq \Xi_B\times \Z^2\times \Z^2$, we identify each composable pair $
\big((\xi,n),(\tau_{-n}(\xi),m)\big)
$
with the triple $(\xi,n,m)$, where $\xi\in\Xi_B$
and $n,m\in\Z^2$.

\medskip

The magnetic field configuration $B$ defines a continuous normalized $2$-cocycle on
$\G_B$. Let us first recall this notion for more general groupoids. 
\begin{definition}
    Let $\G$ be a topological groupoid, and $\G^{(2)}$ its space of composable pairs. A continuous normalized
    $\T$-valued $2$-cocycle on $\G$ is a continuous map
    $\Theta\colon \G^{(2)}\to\T$ satisfying
    $$
        \Theta(g,h)\Theta(gh,k)
        \;=\;
        \Theta(h,k)\Theta(g,hk)$$
    for every composable triple $(g,h,k)$, and
        $\Theta(u,g)=\Theta(g,v)=1$  whenever $u,v\in\G^{(0)}$ are units such that
    $ug=g=gv$.
\end{definition}

 Consider the subgroupoid
$\G_v=\Xi_v\rtimes_{\tau}\Z^2$ of $\G_B$. For $r=(r_1,r_2),\;s=(s_1,s_2)\in \Z^2$  we define the rectangle
$$
\Lambda(r,s)
\;:=\;
\big[\min(r_1,r_1+s_1),\,\max(r_1,r_1+s_1)-1\big]
\times
\big[\min(0,r_2),\,\max(0,r_2)-1\big]
$$
It is not difficult to see that $\Lambda(r,s)$ contains $|s_1r_2|$ points of $\Z^2$
and by convention is empty precisely when $s_1=0$ or $r_2=0$. We define the map $\Theta_v\colon\G_v^{(2)}\to\T$ by setting
$$
\Theta_v(\xi,r,s)
\;:=\;
\prod_{n\in\Lambda(r,s)\cap \Z^2}
e^{-\ii\operatorname{sign}(r_2)\operatorname{sign}(s_1)\xi(-n)},
$$
with the empty product equal to $1$, that is, 
$\Theta_v(\xi,r,s)=1$ if $r_2=0$ or $s_1=0$. 

\begin{proposition}
For every $v\in V$, the map
$\Theta_v\colon\G_v^{(2)}\to\T$ is a continuous normalized
$2$-cocycle.
\end{proposition}

\begin{proof}
Notice that if either $r$ or $s$ is  $0$ the rectangle $\Lambda(r,s)$ is empty. Thus, from the empty-product convention one has that
$\Theta_v(\xi,r,0)=\Theta_v(\xi,0,s)=1$ for all $r,s\in \Z^2$ and $\xi \in \Xi_v.$ This verifies that $\Theta_v$ is normalized.  

For fixed $r,s\in\Z^2$, the value $\Theta_v(\xi,r,s)$
depends on only finitely many coordinates of $\xi$. Hence it
is continuous in $\xi$, and therefore continuous on
$\G_v^{(2)}$, since $\Z^2$ is discrete. To verify the cocycle identity, we first consider the case in which all coordinates of
$r,s,t$ are positive. Throughout the proof, all the
rectangles are understood as subsets of $\Z^2$. Cut the rectangle $\Lambda(r+s,t)$ at the height of $r_2$ into the disjoint
rectangles
$$
A\;:=\;[r_1+s_1,r_1+s_1+t_1)\times[0,r_2),
\qquad
B\;:=\;[r_1+s_1,r_1+s_1+t_1)\times[r_2,r_2+s_2).$$
The lower rectangle $A$ joins $\Lambda(r,s)$ to form
$\Lambda(r,s+t)$, whereas the upper rectangle satisfies
$B=r+\Lambda(s,t)$. Thus,
$$
\Lambda(r+s,t)\;=\;A\sqcup B,
\qquad
\Lambda(r,s)\sqcup A\;=\;\Lambda(r,s+t).$$
Consequently,
\begin{align*}
&\Theta_v(\xi,r,s)\Theta_v(\xi,r+s,t)\\
&\quad=\;
\left(\prod_{n\in\Lambda(r,s)}e^{-\ii\xi(-n)}\right)
\left(\prod_{n\in A}e^{-\ii\xi(-n)}\right)
\left(\prod_{n\in B}e^{-\ii\xi(-n)}\right)\\
&\quad=\;
\Theta_v(\xi,r,s+t)
\prod_{m\in\Lambda(s,t)}e^{-\ii\xi(-(m+r))}\\
&\quad=\;
\Theta_v(\xi,r,s+t)\Theta_v(\tau_{-r}\xi,s,t),
\end{align*}
where the last equality follows from
$\xi(-(m+r))=(\tau_{-r}\xi)(-m)$. For arbitrary signs, the same cutting and joining argument
holds with signed multiplicities. To make this precise,
define the signed interval indicator
$J_{a,b}(k)=\mathbf{1}_{\{k<b\}}-\mathbf{1}_{\{k<a\}}$. We have $
J_{a,c}=J_{a,b}+J_{b,c}$ for all integers $a,b,c$, independently of their order.
Define the corresponding signed rectangle indicator by
$$
R(a,b;c,d)(n_1,n_2)
\;:=\;J_{a,b}(n_1)J_{c,d}(n_2).$$
The signed indicator associated with $\Lambda(r,s)$ is
$$
M_{r,s}
\;:=\;R(r_1,r_1+s_1;0,r_2)
\;=\;
\operatorname{sign}(r_2)\operatorname{sign}(s_1)
\mathbf{1}_{\Lambda(r,s)}.$$
Splitting the second rectangle vertically, exactly as in
the positive case, gives
\begin{align*}
M_{r,s}+M_{r+s,t}
\;&=\;
R(r_1,r_1+s_1;0,r_2)+
R(r_1+s_1,r_1+s_1+t_1;0,r_2)\\
&\;\;\;+R(r_1+s_1,r_1+s_1+t_1;r_2,r_2+s_2).
\end{align*}
Joining the first two terms horizontally therefore yields
$$
M_{r,s}(n)+M_{r+s,t}(n)
\;=\;
M_{r,s+t}(n)+M_{s,t}(n-r).$$
This identity includes all cancellations on overlapping
regions, as well as the cases of empty rectangles. Using these signed multiplicities in the defining products,
we conclude that
\begin{align*}
&\Theta_v(\xi,r,s)\Theta_v(\xi,r+s,t)\\
&\quad=\;
\prod_{n\in\Z^2}
e^{-\ii[M_{r,s}(n)+M_{r+s,t}(n)]\xi(-n)}\\
&\quad=\;
\left(\prod_{n\in\Z^2}
e^{-\ii M_{r,s+t}(n)\xi(-n)}\right)
\left(\prod_{n\in\Z^2}
e^{-\ii M_{s,t}(n-r)\xi(-n)}\right)\\
&\quad=\;
\Theta_v(\xi,r,s+t)
\prod_{m\in\Z^2}
e^{-\ii M_{s,t}(m)(\tau_{-r}\xi)(-m)}\\
&\quad=\;
\Theta_v(\tau_{-r}\xi,s,t)\Theta_v(\xi,r,s+t).
\end{align*}
All products have only finitely many nontrivial factors.
This proves the $2$-cocycle identity.
\end{proof}

Since the same formula is used for every $v\in V$, the cocycles
$\Theta_v$ agree on the overlaps $\Xi_v\cap\Xi_w$. Hence, by the pasting Lemma they define
a continuous normalized $2$-cocycle $\Theta_B$ on $\G_B$ such that
$\Theta_B|_{\G_v^{(2)}}=\Theta_v$.
\begin{proposition}
There exists a unique continuous normalized $2$-cocycle
$\Theta_B\colon\G_B^{(2)}\to\T$ such that
$$
\Theta_B|_{\G_v^{(2)}}\;=\;\Theta_v,
\qquad \forall v\in V.$$
\end{proposition}

\section{The algebra of observables}
In this section, we construct the $C^*$-algebras associated with the different geometric strata of the system. For more information about magnetic algebras we refer to \cite{DGO2026,DG2021,PSbook, Ojito2023}. 
\subsection{The graph magnetic algebra}
The magnetic hull $\Xi_B$ encodes the bulk, edge, and vertex
configurations of the system, while the $\Z^2$-action
and the cocycle $\Theta_B$ describe their spatial covariance and magnetic phases. Combining these data yields the full
magnetic algebra of observables associated with $(\mathfrak G,B)$:
\begin{definition}
    The algebra of the system $(\mathfrak G,B)$ is the \emph{twisted crossed-product} algebra
    $$\A_B\;=\; C(\Xi_B)\rtimes_{\tau,\theta_B}\Z^2\;\simeq\; C^*_r(\G_B,\Theta_B)$$
    where $\theta_B\colon \Z^2\times \Z^2\to C(\Xi_B,\T)$ is given by $\theta_B(n,m)=\Theta_B(\cdot, n,m)$, and the isomorphism is the canonical identification with the twisted groupoid $C^*$-algebra.
\end{definition}
For background on twisted crossed products and groupoid $C^*$-algebras, we refer to \cite{PR1989,Wil} and \cite{Renault1980,Wil2}, respectively. In the crossed-product presentation, formally any element $A\in \A_B$ admits a Fourier decomposition
$$A\;=\;\sum_{n\in \Z^2} a_n U_n$$
where $a_n\in C(\Xi_B)$ and  $n\mapsto U_n$ is projective representation of $\Z^2$ in the sense
$$U_nU_m\;=\; \theta_B(n,m) U_{n+m}$$
Any configuration $\xi\in \Xi_B$ produces a left-regular representation $\pi_\xi\colon \A_B\to \mathscr{B}(\ell^2(\Z^2))$ determined by the matrix elements
\begin{equation}\label{eq: left regular}
    \langle n|\pi_\xi(A)|m\rangle
    \;=\;
    a_{n-m}(\tau_n\xi)\,
    \Theta_B(\tau_n\xi,n-m,m),
    \qquad n,m\in\Z^2.
\end{equation}
According to Proposition \ref{prop: desciprtion of the hull}, up to unitary equivalence, there are three different type of representations $\pi_k\equiv \pi_{b_k}$ where $k\in E\cup F\cup V.$ Our idea now is to describe these representations and their physical meaning.

\medskip

For each $k=f,e,v$, the configuration
$b_k\colon\Z^2\to\R$ determines a vector potential
$A_k\colon\Z^2\times\Z^2\to\R$.
We choose a gauge in which the potential vanishes on vertical
bonds and is defined on horizontal bonds by
\begin{equation*}
    A_k(n,n-e_j)
    \;:=\;
    \delta_{j,1}
    \left(
    \mathbf{1}_{\{n_2>0\}}\sum_{m=1}^{n_2}b_k(n_1,m)
    -
    \mathbf{1}_{\{n_2<0\}}\sum_{m=0}^{|n_2|-1}b_k(n_1,-m)
    \right),
\end{equation*}
where $n=(n_1,n_2)\in\Z^2$ and $j=1,2$.
The potential on reversed bonds is defined by antisymmetry,
and it is set equal to zero on pairs of sites that are not
nearest neighbors. This choice satisfies
$$
A_k(n,n-e_1)
-
A_k(n-e_2,n-e_2-e_1)
\;=\;
b_k(n),
$$
so its circulation around the corresponding oriented
elementary plaquette equals the prescribed magnetic flux. The associated \emph{magnetic translations} in the gauge $A_k$ are the unitary
operators on $\ell^2(\Z^2)$ given by
\begin{equation}\label{eq:mag_tras}
\begin{split}
    (S_{1,k}\psi)(n)
    \;&:=\;
    \expo{\ii A_k(n,n-e_1)}\psi(n-e_1),\\
    (S_2\psi)(n)
    \;&:=\;
    \psi(n-e_2).
\end{split}
\end{equation}
Notice that $S_2$ is independent of $k$ in this gauge. Let $f_k\colon\Z^2\to\T$ be the exponentiated magnetic
flux $
f_k(n):=\expo{\ii b_k(n)},
$
and denote by $M_k$ the corresponding multiplication operator, \ie 
$$
(M_k\psi)(n)
\;:=\;
f_k(n)\psi(n),
\qquad \psi\in\ell^2(\Z^2).
$$
The circulation identity above yields the magnetic
commutation relation
\begin{equation}\label{equ 2.4}
    S_{1,k}S_2
    \;=\;
    M_kS_2S_{1,k}.
\end{equation}

\begin{definition}\cite{DG2021}
For each face, edge, or vertex $k$, the magnetic algebra
$\A_{b_k}$ is the $C^*$-algebra generated by the associated
magnetic translations:
$$
\A_{b_k}
\;:=\;
C^*\big\{S_{1,k},S_2\big\}
\;\subset\;
\mathscr{B}(\ell^2(\Z^2)).
$$
\end{definition}

The magnetic algebra $\A_{b_k}$ is determined, up to unitary equivalence,
by the field $b_k$. Indeed, any two vector potentials with the same
magnetic field differ by a gauge transformation~\cite[Proposition~1]{Deni3}.
As the following proposition shows, the corresponding diagonal unitary
intertwines the magnetic translations and hence conjugates one concrete
realization of $\A_{b_k}$ onto the other.

\begin{proposition}\label{prop: unitary equivalence}
Let $A_k$ and $A_k'$ be two vector potentials for the same magnetic
field $b_k$. There exists a diagonal unitary $U_k$ on $\ell^2(\Z^2)$
such that the associated magnetic translations satisfy
$$
    U_k S_{j,k}^{A} U_k^* \;=\; S_{j,k}^{A'}, \qquad j=1,2.$$
In particular, the two potentials define unitarily equivalent
magnetic algebras.
\end{proposition}

For a face $f$,
the uniform field $b_f\equiv B(f)$ gives rise to the noncommutative
torus $\A_{b_f}$ with parameter $B(f)/(2\pi)$ modulo $\Z$. For an
edge $e$, the field $b_e$ takes a constant value on each side of
a straight interface, and $\A_{b_e}$ describes the corresponding
discrete Iwatsuka-type system \cite{DGO2026,DG2021}. At a vertex
$v$, the field $b_v$ is constant in each quadrant of $\Z^2$, and
$\A_{b_v}$ describes the junction of the four incident bulk regions.  As the following proposition shows, the algebras $\A_{b_k}$ describe the local magnetic systems in the configuration $b_k$, that is,  each
$\A_{b_k}$ is the image of $\A_B$ under a representation unitarily
equivalent to the regular representation at $b_k$.

\begin{proposition}
For $k=f,e,v$, there exists a representation $\rho_k$,
unitarily equivalent to the regular representation $\pi_k$,
such that
$$
    \A_{b_k}\;=\;\rho_k(\A_B).
$$
\end{proposition}

\begin{proof}
Let $\pi_k$ be the regular representation associated with $b_k$
given by \eqref{eq: left regular}. Let $U_n$, $n\in\Z^2$,
denote the canonical unitaries of the twisted crossed product,
and set $T_{j,k}:=\pi_k(U_{e_j})$ with $e_j$ the canonical basis of $\Z^2$. A computation provides that these operators in $\ell^2(\Z^2)$ are the twisted translations
$$
(T_{j,k}\psi)(n)
\;=\;
\Theta_B(\tau_n b_k,e_j,n-e_j)\psi(n-e_j),
$$
and, for $a\in C(\Xi_B)$,
$$
(\pi_k(a)\psi)(n)\;=\;a(\tau_n b_k)\psi(n).
$$
Define $q\in C(\Xi_B)$ by $q(\xi):=e^{\ii\xi(0)}$.
The definition of the magnetic cocycle gives
$$
\Theta_B(\xi,e_1,e_2)\;=\;1,
\qquad
\Theta_B(\xi,e_2,e_1)\;=\;q(\xi)^*.
$$
Thus, the twisted multiplication yields
$$
T_{1,k}T_{2,k}\;=\;\pi_k(q)T_{2,k}T_{1,k},
$$
and consequently
$$
\pi_k(q)
\;=\;
T_{1,k}T_{2,k}T_{1,k}^*T_{2,k}^*
\in C^*(T_{1,k},T_{2,k}).
$$
Since $(\tau_n b_k)(0)=b_k(n)$, we have
$$
(\pi_k(q)\psi)(n)\;=\;e^{\ii b_k(n)}\psi(n).
$$
Hence these translations have the prescribed plaquette phases. By covariance, all represented translates of $q$ also belong
to $C^*(T_{1,k},T_{2,k})$. Let $\Xi_k$ denote the orbit closure
of $b_k$. Under our assumption that the exponentiated field
values are pairwise distinct, the functions $
\xi\mapsto e^{\ii\xi(n)},
$ with $n\in \Z^2$ separate points of $\Xi_k$. By the Stone--Weierstrass theorem,
they generate $C(\Xi_k)$ as a unital $C^*$-algebra.
Since $\pi_k(a)$ depends only on $a|_{\Xi_k}$, it follows that
$$
\pi_k(C(\Xi_B))\;\subset\; C^*(T_{1,k},T_{2,k}).
$$
As $\A_B$ is generated by $C(\Xi_B)$ and $U_{e_1},U_{e_2}$,
we conclude that $
\pi_k(\A_B)=C^*(T_{1,k},T_{2,k}).
$ Thus, the representation $\rho_k$ is obtained by conjugating $\pi_k$ with the unitary provided in Proposition \ref{prop: unitary equivalence}.
\end{proof}

\subsection{The bulk algebra}
Recall that $\mathscr{O}_F$ is a closed invariant subset of
$\Xi_B$ consisting of the constant-field bulk configurations.
We define the \emph{bulk groupoid} as the reduction
$\G_F:=\G_B|_{\mathscr{O}_F}$.
The associated twisted groupoid $C^*$-algebra,
$$
\A_F
\;:=\;
C(\mathscr{O}_F)\rtimes_{\tau,\theta_B}\Z^2\;\simeq\;C^*(\G_F,\Theta_B|_{\G_F^{(2)}}),
$$
is called the \emph{bulk algebra}, as it describes the
constant-field bulk phases of the system.

\begin{proposition}
The bulk algebra admits the decomposition
$$
\A_F\;\simeq\;\bigoplus_{f\in F}\A_{f},
$$
where $\A_{f}$ is the noncommutative torus with parameter
$\theta_f=B(f)/(2\pi)$, understood modulo $\Z$.
\end{proposition}
\begin{proof}
Each $b_f$ is fixed by the translation action. Hence
$
\G_F=\bigsqcup_{f\in F}\big(\{b_f\}\times\Z^2\big)
$ is a finite disjoint union of clopen subgroupoids. Therefore,
$$
\A_F
\;\simeq\;
\bigoplus_{f\in F}
C^*\big(\Z^2,\Theta_B(b_f,\cdot,\cdot)\big)\;=\;\bigoplus_{f\in F} \A_{f}.
$$
For each $f$, the restricted cocycle is $
\Theta_B(b_f,r,s)=e^{-\ii B(f)r_2s_1}.$ The corresponding twisted group algebra is therefore generated
by two unitaries $U_1,U_2$ satisfying
$$
U_1U_2\;=\;e^{\ii B(f)}U_2U_1.
$$
It is thus the noncommutative torus with parameter
$\theta_f=B(f)/(2\pi)$.
\end{proof}

The $K$-theory of $\A_F$ is well known
\cite{gracia-varilly-figueroa-01,Con}:
$$
K_0(\A_F)\;\simeq\;\Z^{2|F|}
\;\simeq\;K_1(\A_F).
$$
For each $f\in F$, the group $K_0(\A_f)$ is freely generated by the
identity class $[{\bf 1}_f]$ and a class $[P_{b_f}]$. If
$B(f)\notin 2\pi\Z$, we take $P_{b_f}$ to be a Powers--Rieffel
projection~\cite{gracia-varilly-figueroa-01}. If $B(f)\in 2\pi\Z$,
then $\A_f\simeq C(\T^2)$, and we take $P_{b_f}$ to be a Bott
projection in ${\rm M}_2(C(\T^2))$. While
$K_1(\A_{b_f})$ is generated by the classes
$[U_{1,f}]$ and $[U_{2,f}]$ of the canonical
unitaries of the noncommutative torus.
Consequently, 
$$
K_0(\A_F)
\;=\;
\bigoplus_{f\in F}
\big(\Z[{\bf 1}_f]\oplus\Z[P_{b_f}]\big), \qquad 
K_1(\A_F)
\;=\;
\bigoplus_{f\in F}
\big(\Z[U_{1,f}]\oplus\Z[U_{2,f}]\big).
$$

\subsection{Interface algebra}
The magnetic interfaces of the system contains the edges and the vertices of the graph. This information is encoded in the open invariant subset $\mathscr{O}_{V\cup E}:=\mathscr{O}_V\sqcup \mathscr{O}_E$ of $\Xi_B.$ The reduction of $\A_B$ to $\mathscr{O}_{V\cup E}$ produces the interface algebra
\begin{equation}
  \A_{V\cup E}\;:=\;C_0(\mathscr{O}_{V\cup E})\rtimes_{\tau,\theta_B}\Z^2
\end{equation}
Notice that $\A_{V\cup E}$ is aclosed two-sided ideal of $\A_B$ and since $\mathscr{O}_{V\cup E}=\Xi_B\setminus \mathscr{O}_F$, one gets the exact sequence
\begin{equation}\label{eq: bulk-interface sequence}
    0\to \A_{V\cup E}\to \A_B\to \A_{F}\to 0
\end{equation}
Two important invariant subsets of $\mathscr{O}_{V\cup E}$ are the open subset $\mathscr{O}_V$ and the closed subset $\mathscr{O}_E$. By restriction, one gets the \emph{edge} and \emph{vertex} algebra defined as
$$\A_E\;=\; C_0(\mathscr{O}_E)\rtimes_{\tau,\theta_B}\Z^2,\qquad \A_V\;=\;C_0(\mathscr{O}_V)\rtimes_{\tau,\Theta_B} \Z^2$$
In order to get a description of $\A_E$ and $\A_V$ we start with the following technical Lemma.
\begin{lemma}\label{lemm: trivial twist}
For every $\zeta\in\mathscr{O}_{V\cup E}$, the restriction
of $\Theta_B$ to the orbit groupoid
$\G_B|_{\mathscr{O}(\zeta)}$ is cohomologically trivial.
\end{lemma}
\begin{proof}
Set $\mathscr O:=\mathscr O(\zeta)$ and
$\G_{\mathscr O}:=\G_B|_{\mathscr O}$.
It suffices to show that $H^2(\G_{\mathscr O},\T)=0$.
Let $H_{\mathscr O}$ be the isotropy group at $\zeta$.
Then $H_{\mathscr O}=\{0\}$ for a vertex configuration,
whereas $H_{\mathscr O}\simeq\Z$ for an edge configuration.
Moreover, the orbit is discrete and admits the identification
$\mathscr O\simeq\Z^2/H_{\mathscr O}$.

Identifying a continuous groupoid cochain $c(\xi,n_1,\ldots,n_p)$
with the group cochain $(n_1,\ldots,n_p)\mapsto
c(\cdot,n_1,\ldots,n_p)$, the groupoid coboundary
\cite[Definition~4.6]{GillaspyKumjian} agrees with the group
coboundary for the action $(n\cdot f)(\xi)=f(\tau_{-n}\xi)$.
Hence
$$
H^2(\G_{\mathscr O},\T)
\;=\;H^2(\mathscr O\rtimes\Z^2,\T)
\;\simeq\; H^2\big(\Z^2,C(\mathscr O,\T)\big).
$$
Since $\mathscr O$ is discrete, $C(\mathscr O,\T)$ consists
of all $\T$-valued functions on $\mathscr O$. With its natural
$\Z^2$-action, this is the coinduced module
\cite[Definition~6.3.1]{Weibel}
$
C(\mathscr O,\T)
\simeq \operatorname{Coind}_{H_{\mathscr O}}^{\Z^2}\T,$ 
where $H_{\mathscr O}$ acts trivially on $\T$.
Shapiro's lemma \cite[Lemma~6.3.2]{Weibel} therefore yields
$$
H^2(\G_{\mathscr O},\T)\;\simeq\; H^2(H_{\mathscr O},\T)\;=\;0,
$$
since the trivial group has vanishing cohomology in positive
degrees, and $\Z$ has vanishing cohomology in degrees at least
two \cite[Example~6.1.4]{Weibel}.
\end{proof}
\begin{proposition}
    It holds true that
    \begin{equation}
 \A_V\;=\; \bigoplus_{v\in V}\A_{v},\qquad \A_E\;= \; \bigoplus_{e\in E}\A_{e}.
\end{equation}
where $\A_{v}\simeq \mathfrak K(\ell^2(\Z^2))$ and $\A_{e}\simeq \mathfrak K(\ell^2(\Z))\otimes C(\T)$, with $\mathfrak K$ the algebra of compact operators.
\end{proposition}
\begin{proof}
Since the vertex and edge strata are disjoint unions
of invariant open orbits, the crossed products decompose as
$$
\A_V\;\simeq\;\bigoplus_{v\in V}\A_{v},
\qquad
\A_E\;\simeq\;\bigoplus_{e\in E}\A_{e}.
$$
By Lemma~\ref{lemm: trivial twist}, the twist can be
removed on each orbit. Thus
$$
\A_{v}\;\simeq\; C_0(\mathscr O_v)\rtimes_\tau\Z^2,
\qquad
\A_{e}\;\simeq \;C_0(\mathscr O_e)\rtimes_\tau\Z^2.$$
For a vertex configuration, $\mathscr O_v\simeq\Z^2$
equivariantly. Since $\Z^2$ acts on itself by left translation,
\cite[Theorem~4.24]{Wil} gives
$$
\A_{v}\;\simeq\; C_0(\Z^2)\rtimes\Z^2
\;\simeq\;\mathfrak K(\ell^2(\Z^2)).$$
For an edge configuration, write
$\Z^2=\Z_\perp\oplus\Z_\parallel$, where
$\Z_\parallel$ is the isotropy subgroup.
Under the identification $\mathscr O_e\simeq\Z$,
$\Z_\perp$ acts by translations and
$\Z_\parallel$ acts trivially. Consequently,
$$
\begin{aligned}
\A_{e}\;\simeq\;
\big(C_0(\Z)\rtimes\Z_\perp\big)
\rtimes_{\mathrm{id}}\Z_\parallel
\;\simeq\;
\mathfrak K(\ell^2(\Z))\otimes C^*(\Z)
\;\simeq\;
\mathfrak K(\ell^2(\Z))\otimes C(\T).
\end{aligned}
$$
This completes the proof.
\end{proof}

It follows from the construction that the interface algebra enters in the following exact sequence
\begin{equation}\label{eq: exact sequence edge-corner}
    0\to \A_V\to \A_{V\cup E}\to\A_E\to 0
\end{equation}

Let us notice that
$$
K_0(\A_V)\;=\;\bigoplus_{v\in V}\Z[P_v],
\qquad K_1(\A_V)\;=\;0,$$
where $P_v$ corresponds to the rank-one projection
$|\delta_0\rangle\langle\delta_0|$ on $\ell^2(\Z^2)$.
Similarly,
$$
K_0(\A_E)\;=\;\bigoplus_{e\in E}\Z[P_e],
\qquad
K_1(\A_E)\;=\;\bigoplus_{e\in E}\Z[S_e].
$$
Here $P_e$ corresponds to
$|\delta_0\rangle\langle\delta_0|\otimes1$ under
$\A_e\simeq\mathfrak K(\ell^2(\Z))\otimes C(\T)$,
while, under the stability isomorphism, $[S_e]$
corresponds to the class of the bilateral shift in $\ell^2(\Z)$
implementing the forward translation along the oriented edge $e$. Therefore, the unique non-trivial boundary map associated with the six-term exact
sequence of \eqref{eq: exact sequence edge-corner} is 
$
\partial_1^{\rm EV}\colon
K_1(\A_E)\to K_{0}(\A_V).$
The Toeplitz index map gives opposite signs at the initial and terminal endpoints, yielding
$$
\partial_1^{\rm EV}([S_e])
\;=\;[P_{\partial_1(e)}]-[P_{\partial_0(e)}],$$
where recall that $\partial_0(e)$ and $\partial_1(e)$ denote
the initial and final vertices of $e$, respectively.
For the rectangular graphs considered here, ${\rm Ker}(\partial_1^{\rm EV})$ is generated by the oriented face classes
$
[S_f]:=\sum_{e\in\partial f}{\rm sign}(f,e)[S_e]$. Their unique relation is $
\sum_{f\in F}[S_f]=0,$
because each edge belongs to exactly two faces and
occurs with opposite orientations. Since $K_1(\A_V)=0$, exactness gives $
K_1(\A_{V\cup E})
\simeq{\rm Ker}(\partial_1^{\rm EV})$
and
$$
0\longrightarrow
\operatorname{Coker}(\partial_1^{\rm EV})
\longrightarrow K_0(\A_{V\cup E})
\longrightarrow K_0(\A_E)
\longrightarrow0.$$
For an unbounded edge, the incidence formula contains only the
term corresponding to its finite endpoint. Thus the image of
$\partial_1^{\rm EV}$ contains a vertex generator. Since the
finite edges connect all vertices, their endpoint differences
generate all remaining vertex classes together with this generator. Hence $\partial_1^{\rm EV}$ is surjective, and exactness gives
$K_0(\A_{V\cup E})\simeq K_0(\A_E)$.

\begin{proposition}
For the connected rectangular graphs considered above,
$$
K_1(\A_{V\cup E})\;\simeq\;\Z^{|F|-1},\quad K_0(\A_{V\cup E})\;\simeq\;\Z^{|E|}
$$
\end{proposition}
\subsection{The cofiltration by stratum}
The invariant subset
$\mathscr O_{F\cup E}:=\mathscr O_F\cup\mathscr O_E$
describes the asymptotic configurations of the system
away from the vertices. We define the associated
asymptotic algebra by
$$
\A_{F\cup E}\;:=\;
C_0(\mathscr O_{F\cup E})
\rtimes_{\tau,\theta_B}\Z^2.$$
It encodes the bulk regimes within the faces and the
interface regimes along the edges. Thanks to Proposition \ref{prop: desciprtion of the hull}, one gets the following cofiltration of the $\A_B$
\begin{equation}\label{eq: cofiltration}
    \A_B\equiv\A_{F\cup E\cup V}
    \xrightarrow{\varphi_{F\cup E}}
    \A_{F\cup E}
    \xrightarrow{\varphi_F}
    \A_F
\end{equation}
It follows from the construction that
${\rm Ker}(\varphi_{F\cup E})= \A_V$ and ${\rm Ker}(\varphi_{F})= \A_E$. The stratification of the magnetic hull yields the short exact sequence
\begin{equation}\label{eq: bulk-edge sequence}
0\longrightarrow \A_E
\longrightarrow \A_{F\cup E}
\xrightarrow{\varphi_F}\A_F
\longrightarrow 0.
\end{equation}
Its connecting maps in $K$-theory relate bulk observables on the
faces to interface observables along the edges and define the
\emph{bulk--edge correspondence} \cite{PSbook}. Similarly, the
connecting maps of \eqref{eq: exact sequence edge-corner} relate
edge observables to corner observables at the vertices and define
the \emph{edge--corner correspondence} \cite{Hay, Ojito2023}. These connecting maps are
$$
\partial_i^{\rm FE}\colon
K_i(\A_F)\longrightarrow K_{1-i}(\A_E),
\qquad
\partial_i^{\rm EV}\colon
K_i(\A_E)\longrightarrow K_{1-i}(\A_V),
\qquad i=0,1.$$

\section{Topological currents}
\subsection{Traces on the observable algebras}
The crossed product structure of $\A_B$ yields a canonical faithful
conditional expectation $E\colon\A_B\to C(\Xi_B)$, given on finite
Fourier sums by $E(\sum_n a_nU_n)=a_0$ \cite{Dav}. Since $\Xi_B$ is
compact, each positive $\Z^2$-invariant Radon measure $\mu$ on $\Xi_B$
defines a bounded trace on $\A_B$ via
$$
\mathcal T_\mu(A)\;:=\;\int_{\Xi_B}E(A)(\xi)\,d\mu(\xi).
$$
This trace is faithful if $\mu$ has full support and is a tracial state
if $\mu$ is a probability measure.  For $K=F,E,V$, the same formula, with $E$ replaced by the canonical
expectation $E_K\colon\A_K\to C_0(\mathscr O_K)$ and integration
taken over $\mathscr O_K$, associates to each positive invariant Radon
measure on $\mathscr O_K$ a densely defined lower semicontinuous
tracial weight on $(\A_K)_+$. Thus, restricting $\mu\in\mathfrak M$
to each stratum yields the corresponding tracial weight. In particular,
the invariant measures $\mu_f$, $\mu_e$, and $\mu_v$ from
Proposition~\ref{prop: measures} yield the bulk, edge, and vertex
traces $\mathcal T_f$, $\mathcal T_e$, and $\mathcal T_v$,
respectively.

\subsection{Numerical Invariants and their correspondences}
 For each $f\in F$, let $\nabla_1$ and
$\nabla_2$ be the commuting $*$-derivations generating the canonical
torus $\T^2$-action on $\A_f$, normalized by $
    \nabla_i(U_j)=2\pi\ii\,\delta_{ij}U_j$
for the canonical unitaries $U_1,U_2$ generating $\A_{B(f)}$ \cite[Chap.~IV, \S6.$\alpha$]{Con}.
An element is \emph{smooth} if its orbit map under this $\T^2$-action is
$C^\infty$. The same dual $\T^2$-action is defined on each of the other observable algebras through its crossed-product presentation. For a bulk projection $P=(P_f)_{f\in F}\in\A_{F}$ with smooth
components, we define its \emph{Chern number} at the face $f$ by
\begin{equation}
    {\rm Ch}(P_f)\;=\;
    \frac{1}{2\pi\ii}\mathcal{T}_f
    \big(P_f[\nabla_1 P_f,\nabla_2 P_f]\big).
\end{equation}
With this normalization, ${\rm Ch}(P_f)$ is an integer and depends
only on the $K_0$-class of $P_f$ \cite[Chap.~III, \S3, Prop.~2(a) and Cor.~16]{Con}.

\medskip

Let $\nabla_e$ denote the $*$-derivation generating the dual action
in the direction of the edge $e$, and write $L^1(\A_e)$ for the
$L^1$-space associated with the edge trace $\mathcal T_e$. For
$k\in\mathbb N$, set
$$
\mathcal C^k_e\;:=\;\big\{\,A\in\mathfrak A_e\mid
\nabla_e^k(A)\in L^1(\A_e)\,\big\}.
$$
Let $U\in\A_e^\sim$ be a unitary such that
$U-{\bf 1}\in\mathcal C^1_e$. The \emph{noncommutative winding numbers} of $U$ is defined as 
\begin{equation}
    \mathcal{W}_{e}(U)\;:=\;\ii\mathcal{T}_e(U^*\nabla_{e}(U))\;.
\end{equation}
The quantity $\W_e(U)$ is constant on the equivalence class of $U$ in $K_1(\A_e)$ \cite[Chap.~III, \S3, Prop.~3(a)]{Con}. The total winding number is the homomorphism $\W\colon K_1(\A_E)\to \Z$  given by the contribution in all the edges, \ie $\mathcal{W}:=\sum_e\W_e.$
\begin{proposition}\label{prop: bulk-edge}
    Let $\partial_i^{\rm 
    FE}\colon K_i(\A_F)\to K_{1-i}(\A_E)$ be the boundary map related to \eqref{eq: bulk-edge sequence}. For any $[P]=([P_f])_{f\in F}\in K_0(\A_F)$ with smooth components it holds that
    $$\W\big(\partial_0^{\rm FE}([P])\big)\;=\;\sum_{f\in F}\sum_{e\in \partial f}{\rm sign}(f,e){\rm Ch}([P_f])$$
    In particular,
    $$\W\big(\partial_0^{\rm FE}([P_f])\big)\;=\;\sum_{e\in \partial f}{\rm sign}(f,e){\rm Ch}([P_f]),\qquad \forall f\in F$$
    
\end{proposition}
\begin{proof}
Any projection $[P_f]\in K_0(\A_f)$ admits the decomposition
$$[P_f]\;=\;a_f[{\bf 1}_f]+{\rm Ch}([P_f])[P_{b_f}],\qquad a_f\in \Z$$
Since $ \partial_0^{\rm FE}([{\bf 1}_f])=0$, one gets
\begin{align*}
    \partial_0^{\rm FE}([P])\;=\; \sum_{f\in F}{\rm Ch}([P_f])\,\partial_0^{\rm FE}([P_{b_f}]).
\end{align*}
By naturality, the $e$-component of the boundary map is computed
in the corresponding edge extension. For the
projection $P_{b_f}$, this component vanishes if $e\notin\partial f$
and equals ${\rm sign}(f,e)[S_e]$ if $e\in\partial f$, by the
Toeplitz bulk--edge correspondence and the chosen edge orientation.
Hence $$
    \partial_0^{\rm FE}([P_{b_f}])
    \;=\;\sum_{e\in\partial f}{\rm sign}(f,e)[S_e].$$
Finally, since $\W_e([S_e])=1$ for each $e\in E$, applying
$\W=\sum_{e\in E}\W_e$ to $\partial_0^{\rm FE}([P])$ gives
the stated formula.
\end{proof}

The first important observation in order to get  topological Kirchhoff law is the following relation between the boundary maps.
\begin{proposition}\label{prop: boundary composition}
       The boundary maps satisfy $$\partial_{1-i}^{\rm EV}\circ \partial_i^{\rm FE}\;=\;0,\qquad i=0,1$$
\end{proposition}
\begin{proof}
   The case $i=1$ is a direct consequence of that $K_1(\A_V)=0.$  Now let us come to the case $i=0$. From the proof of Proposition \ref{prop: bulk-edge} one knows that
    ${\rm Im}(\partial_0^{\rm FE})$ is a $\Z$-module generated by the elements $[S_f]\in K_1(\A_E)$ with $f\in F$ defined as
    $$[S_f]\;=\; \sum_{e\in \partial f}{\rm sign}(f,e)[S_e]$$
On other hand, $\partial^{\rm EV}_1([S_e])=[P_{\partial_1(e)}]-[P_{\partial_0(e)}]$. Then, 
$$\partial^{\rm EV}_1([S_f])\;=\;\sum_{e\in \partial f}{\rm sign}(f,e)\big([P_{\partial_1(e)}]-[P_{\partial_0(e)}]\big)\;=\;0$$
where we used that the contributions at each vertex of $\partial f$ cancel. This verifies that $\partial_{1}^{\rm EV}\circ \partial_0^{\rm FE}=0.$

\end{proof}

\subsection{Hall conductivity}
The system in the faces is modeled by a family of bulk-gap Hamiltonian
$H\in \A_{F}\otimes {\rm M}_N(\C)$, \ie a selfadjoint element $H=(H_f)_{f\in F}$
for which there exists a compact interval $\Delta\subset\R$
with nonempty interior such that
${\rm Spec}(H_f)\cap\Delta=\emptyset$ for all $f\in F$ and
$$
\min {\rm Spec}(H_f)\;< \; \min \Delta\;<\;\max \Delta\;< \; \max {\rm Spec}(H_f)$$
Thus, $\Delta$ lies in a non-trivial common spectral gap of all bulk components. The matrix factor ${\rm M}_N(\C)$ accounts for $N$ internal
degrees of freedom at each lattice site, such as spin or orbital
degrees of freedom.
For any Fermi level $\mu\in \Delta$ one can consider  the \emph{Fermi projection} $$P_\mu\;=\;(P_{\mu,f})_{f\in F}\in \A_F\otimes {\rm M}_N(\C),\qquad P_{\mu,f}\;:=\;\chi_{(-\infty,\mu]}(H_f)$$ 
By stabilization of the K-theory, it defines a class $[P_\mu]\in K_0(\A_F)$.  Henceforth, we suppress the matrix factor ${\rm M}_N(\C)$ in the
notation. We extend maps and derivations entrywise and use the
unnormalized matrix trace ${\rm Tr}_N$ to extend each trace
$\mathcal T$ to $\mathcal T\otimes{\rm Tr}_N$.

\medskip

If $H\in \mathcal{C}^1(\A_{F})$, the \emph{transverse Hall
conductance} of the bulk component associated with the face $f$
is defined by
\begin{equation}
     \sigma_f(\Delta)
    \;=\;
    \frac{\mathfrak e^2}{h}
    {\rm Ch}([P_{\mu,f}]),
    \qquad \mu\in\Delta,
\end{equation}
where recall that $\mathfrak e>0$ is the magnitude of the electron charge, and $h=2\pi\hbar$ is the Planck's constant. Since $\Delta$ lies in a spectral gap of $H_f$, this definition
is independent of the choice of $\mu\in\Delta$.

\subsection{Edge currents}The full Hamiltonian incorporates both the bulk dynamics in each
face and the couplings across the interfaces between neighboring
regions. Accordingly, the full system is modeled by a selfadjoint
lift $\hat H_{\rm total}\in\A_B$ of $H$, \ie a selfadjoint element satisfying
$\varphi_{F}\circ \varphi_{F\cup E}(\hat H_{\rm total})=H$. One can also consider the system information faces and edges via the selfadjoint Hamiltonian $\hat H$ in $\A_{F\cup E}$ given by $\varphi_{F\cup E}(\hat H_{\rm total})=\hat H.$ By construction, one also gets the relation $\varphi_F(\hat H)=H.$

\medskip 

Let $g$ be any non-decreasing smooth function which is $0$  below $\Delta$ and $1$ above. 
Then one can define the unitary operator $U_\Delta:=e^{2\pi \ii g(\hat H)}\in \A_{F\cup E}$. Since it is trivial in the bulk, it defines a class
$[U_\Delta]\in K_1(\A_{ E}).$ We have the following observation.
\begin{proposition}
    Let $\partial_0^{FE} \colon K_0(\A_F)\to K_1(\A_{ E})$. For any $\mu\in \Delta$ it follows that $\partial_0^{\rm FE}([P_\mu])=-[U_\Delta]$ and $$\sum_{\{f\mid e\in \partial f\}}{\rm sign}(f,e){\rm Ch}(P_{\mu,f})\;=\;-\W_e(U_\Delta)$$
\end{proposition}
\begin{proof}
Functional
calculus gives $
    g(H)=1-P_\mu.$ Thus, $1-g(\hat H)$ is a selfadjoint lift of $P_\mu$. By the
definition of the exponential boundary map,
$$    \partial_0([P_\mu])
    \;=\;
[e^{2\pi\ii(1-g(\hat H))}]
    \;=\;
    [U_\Delta^*]
    \;=\;
    -[U_\Delta].$$
    Thus, the remaining part of the proof is a direct consequence of Proposition \ref{prop: bulk-edge}.
\end{proof}

The \emph{edge-current density} carried by the states with energies in
the gap $\Delta$ along an oriented edge $e\in E$ is defined by
\begin{equation}\label{eq: edge current}
J_e(\Delta)
\;:=\;
\frac{\mathfrak e}{\hbar}\,
\mathcal T_e\big(
g'(\widehat H)\nabla_e(\widehat H)
\big),
\end{equation}
Notice that $\operatorname{supp}(g')\subset\Delta$, so that
$g'(\widehat H)$ localizes the observable to the states whose energies
lie inside the bulk gap $\Delta$. The operator $
\frac{1}{\hbar}\nabla_e(\widehat H)$
represents the velocity in the direction of the oriented edge $e$.
Thus, \eqref{eq: edge current} measures the electric current transported along $e$ per unit length.  Under the identification, $\A_e\simeq\mathfrak K(\ell^2(\Z))\otimes C(\T),$ The derivation $\nabla_e$  acts on the longitudinal
variable by
$$
\nabla_e(a\otimes f)
\;=\;
a\otimes\frac{d}{dk}f,
\qquad
a\in\mathfrak K(\ell^2(\Z)),
\quad f\in C^\infty(\T).$$
In particular, $\nabla_e(S_e)=-\ii S_e$.

\begin{proposition}\label{prop: edge currents}
   Assume that
$U_\Delta-{\bf 1}\in \mathcal C^1_e$. It holds true that
    $$J_e(\Delta)\;=\;-\frac{\mathfrak e}{2\pi\hslash}\W_e(U_\Delta)$$
    Consequently, the edge current is proportional to the relative Chern number of the adjacent faces
        $$J_e(\Delta)\;=\;\frac{\mathfrak e}{2\pi\hslash}\sum_{\{f\mid e\in \partial f\}}{\rm sign}(e,f){\rm Ch}(P_{\mu,f})$$
\end{proposition}
\begin{proof}
The proof is based on \cite[Proposition 7.1.2]{PSbook}. By the Duhamel
formula,
$$
\nabla_e(U_\Delta)
\;=\;
2\pi\mathrm i\int_0^1
e^{2\pi\mathrm i(1-t)g(\widehat H)}
\nabla_e\big(g(\widehat H)\big)
e^{2\pi\mathrm itg(\widehat H)}
\,dt.
$$
Multiplying the above identity on the left by $U_\Delta^*$, we use
$\mathcal T_e(\nabla_e(U_\Delta))=0$ in the first equality below
and the edge-current identity of
\cite[Proposition~7.1.2, Eq.~(7.4)]{PSbook} in the second:
\begin{align*}
\mathcal T_e\big(U_\Delta^*\nabla_e(U_\Delta)\big)
\;&=\;\mathcal T_e\big((U_\Delta^*-\mathfrak{1})\nabla_e(U_\Delta)\big)\;=\;2\pi\mathrm i\,
  \mathcal T_e\big(g'(\widehat H)\nabla_e(\widehat H)\big).
\end{align*}
Here $\mathcal T_e(\nabla_e(U_\Delta))=0$ follows from periodicity
in the longitudinal variable and the assumption that
$\nabla_e(U_\Delta)\in L^1(\A_e)$. Therefore,
$
\mathcal W_e(U_\Delta)=
-2\pi
\mathcal T_e\big(
g'(\widehat H)\nabla_e(\widehat H)
\big).
$
It follows from the definition of the edge-current density that
$$
J_e(\Delta)
\;=\;
\frac{\mathfrak e}{\hbar}
\mathcal T_e\big(
g'(\widehat H)\nabla_e(\widehat H)
\big)
\;=\;
-\frac{\mathfrak e}{2\pi\hbar}
\mathcal W_e(U_\Delta).$$
The last assertion follows immediately from Proposition
\ref{prop: bulk-edge}.
\end{proof}
We can now state Kirchhoff's current law.
\begin{theorem}
Let $\widehat H\in \A_{F\cup E}\otimes {\rm M}_N(\C)$ be a smooth selfadjoint element and $\Delta$ lie in a common spectral gap of its bulk Hamiltonians $H=(H_f)_{f\in F}\in \A_F\otimes {\rm M}_N(\C)$. Then, at every vertex $v\in V$, the topological interface current
densities satisfy
$$
    \sum_{\{e\in E \mid \mathfrak \partial_1(e)=v\}} J_e(\Delta)
    \;=\;
    \sum_{\{e\in E\mid \partial_0(e)=v\}} J_e(\Delta).$$
In other words, the total incoming current density equals the total outgoing current density at each vertex.
\end{theorem}
\begin{proof}
Since $[U_\Delta]=-\partial_0^{\rm FE}([P_\mu])$, Proposition~\ref{prop: boundary composition}
gives $\partial_1^{\rm EV}([U_\Delta])=0$. Taking the coefficient of
$[P_v]$ and using $\W_e([S_e])=1$, we obtain
$$
    \sum_{\partial_1(e)=v}\W_e(U_\Delta)
    \;=\;
    \sum_{\partial_0(e)=v}\W_e(U_\Delta).$$
Multiplying by $-\mathfrak e/(2\pi\hbar)$ and applying
Proposition~\ref{prop: edge currents} proves the claim.
\end{proof}

     \end{document}